\documentclass[]{piparticle-final}
\usepackage{graphicx}
\usepackage{amsmath}
\usepackage{epstopdf}

\usepackage{cite}
\begin{document}

\volume{4}               % To be inserted by Editor
\articlenumber{040003}   % To be inserted by Editor
\journalyear{2012}       % To be inserted by Editor
\editor{V. Lakshminarayanan}   % To be inserted by Editor
\reviewers{C. Negreira, Laboratorio de Ac\'ustica Ultrasonora, \\ \mbox{}\hspace{38mm} Universidad de la Rep\'ublica, Uruguay}  % To be inserted by Editor
\received{14 March 2011}     % To be inserted by Editor
\accepted{23 April 2012}   % To be inserted by Editor
\runningauthor{E. Fern\'andez \textit{et al.}}  % To be inserted by Editor
\doi{040003}         % To be inserted by Editor
\title{Natural and laser-induced cavitation in corn stems: On the mechanisms of acoustic emissions}
\author{E. Fern\'andez,\cite{inst1} 
        R. J. Fern\'andez,\cite{inst1}
        G. M. Bilmes\cite{inst2}\thanks{E-mail: gabrielb@ciop.unlp.edu.ar}}

\pipabstract{
Water in plant xylem is often superheated, and therefore in a meta-stable state. Under certain conditions, it may suddenly turn from the liquid to the vapor state. This cavitation process produces acoustic emissions. We report the measurement of ultrasonic acoustic emissions (UAE) produced by natural and induced cavitation in corn stems. We induced cavitation and UAE \textit{in vivo, }in well controlled and reproducible experiments, by irradiating the bare stem of the plants with a continuous-wave laser beam. By tracing the source of UAE, we were able to detect absorption and frequency filtering of the UAE propagating through the stem. This technique allows the unique possibility of studying localized embolism of plant conduits, and thus to test hypotheses on the hydraulic architecture of plants. Based on our results, we postulate that the source of UAE is a transient ``cavity oscillation'' triggered by the disruptive effect of cavitation inception.
}

\maketitle

\blfootnote{
\begin{theaffiliation}{99}
   \institution{inst1} IFEVA, Facultad de Agronom\'{\i}a, Universidad de Buenos Aires y CONICET, Av. San Mart\'{\i}n 4453, C1417DSE Buenos Aires, Argentina.
   \institution{inst2} Centro de Investigaciones Opticas (CONICET-CIC) and Facultad de Ingenier\'{\i}a, Universidad Nacional de La Plata, Casilla de Correo 124, 1900 La Plata, Argentina.
\end{theaffiliation}
}

\section{Introduction}

The cohesion-tension theory suggests that water in the xylem of transpiring plants is under tension with a hydrostatic pressure below atmospheric and, thus, most of the time at ``negative'' values \cite{1}. Negative pressures means that water in the xylem has a reduced density compared to equilibrium \cite{2}. According to its phase diagram, water under these conditions is overheated (i.e., in a meta-stable state). Therefore, it should not be in the liquid but in the vapor phase \cite{3}. The molecules in the liquid phase are further away from each other, but their mutual attraction allows the system to remain unchanged.

Under sufficiently high tension (i.e., low pressures caused by water deficit), xylem may fail to maintain this state, causing liquid water to turn into vapor in a violent way. This phenomenon, usually known as cavitation, causes the embolism of the conduits, reducing tissue hydraulic conductivity and exacerbating plant physiological stress \cite{4,5}. Some herbaceous species are known to sustain cavitation almost every day, repairing embolism during the night, while most woody species preclude cavitation occurrence by a combination of stomatal behavior and anatomical and morphological adjustment \cite{6}. 

Cavitation events in xylem produce sound \cite{7,8}. In 1966, Milburn and Johnson developed a technique to detect sound by registering `clicks' in a record player pick-up head attached to stressed plants and connected to an amplifier \cite{9}. They associated this sound emission with the rupture of the water column in xylem vessels. Since then, several authors have used this audible acoustic emission technique to measure xylem cavitation \cite{10,11}. Later on, some authors have improved the technique by detecting ultrasonic acoustic emissions (UAE) \cite{12,13,14,15,16,17}. These authors have demonstrated a good correlation between UAE and cavitation. However, the connection between audible or ultrasonic acoustic emissions and cavitation phenomena on xylem vessels remains unexplained. Tyree and Dixon \cite{12} proposed four possible sources of acoustic emissions that we will consider in the Discussion section. Others authors have developed explanations based on alternatives to the cohesion-tension theory \cite{3,18}

One of the problems of studying cavitation in plants is the spontaneous character of the phenomenon, so far precluding our ability to produce it in a controlled way. Most cavitation experiments use transpiration to raise xylem water tension, the trigger for cavitation events. In some cases, xylem tension was increased by centrifugation \cite{19, 20}, but even there, cavitation events took place rather randomly along the water column. On the other hand, in order to study bubble behavior in isolated physical systems, several authors explored the generation of cavitation phenomena using lasers \cite{21,22,23,24}. In these experiments, cavitation bubbles are generated in a very well defined location taking advantage of the accuracy of the laser beam. Even though this technique was developed to generate cavitation in transparent environments, we wondered whether it could be used in biological systems to generate cavitation at specific locations along the stem.

In this article, we report spontaneous UAE produced by natural cavitation in xylem vessels of corn (\textit{Zea mays }L.) stems and we characterized and classified the signals. We also developed a method to produce laser-induced cavitation and UAE events in a controlled way by irradiating plants with a continuous-wave laser. We performed experiments with this method to study the generation and propagation of UAE. Our results allowed us to explain the connection between cavitation and UAE, as well as the relationship between signal frequency and the localization of the source in the stem.

\section{Materials and methods}

\begin{figure}[th]
\begin{center}
\includegraphics[width=0.95\columnwidth]{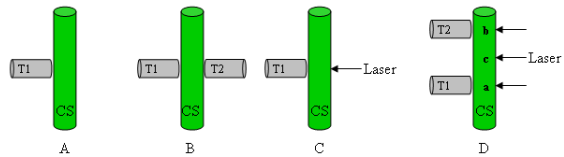}
\caption{Set-up for the different experiments. (A) Experiment 1. (B) Experiments 2 and 4. (C) Experiment 3. (D) Experiment 5. CS: corn stem; T1 and T2: PZT transducers for UAE detection.} \label{figure1}
\end{center}
\end{figure} 

Corn plants were grown in a greenhouse in 3l pots containing sand. They were watered at field capacity every 1-2 days with nutritive solution (3 g~l$^{-1}$ of KSC II -- Roulier). After four months, tasseling plants (around 1 m high and 12 mm stem width) were used to perform experiments under different conditions: total darkness (D); room diffuse light (RL; PAR ca. 100 mE m$^{-2}$ s$^{-1}$); leaf illumination with a 150 W incandescent lamp (IL; placed ca. 0.5 m away), and laser irradiation (L). In the latter case, experiments were carried out by directing the beam of a 50 mW He--Ne red laser (630 nm), or a continuous-wave (CW) Ar-ion laser (Spectra Physics Model 165/09) directly on to the stems. Most experiments were conducted sequentially in 3-5 plants, and we report the full range of observed results.

Ultrasonic acoustic emissions generated in the stems of plants were recorded by home-made PZT-based piezoelectric transducers (4 $\times$ 4 mm, 230 kHz) \cite{25}  coated with glycerin and clamped to the bare stem by a three-prong thumbtack. Signals, of the order of 1 mV, were amplified (gain $10^3$) and recorded in a storage digital oscilloscope. Different transducer positions in the stems were explored, as well as simultaneous measurement of UAE with two detectors attached to different points of the stems, providing a method to trace the origin of the signals (Fig. 1).

\begin{figure*}[th]
\begin{center}
\includegraphics[width=.8\textwidth]{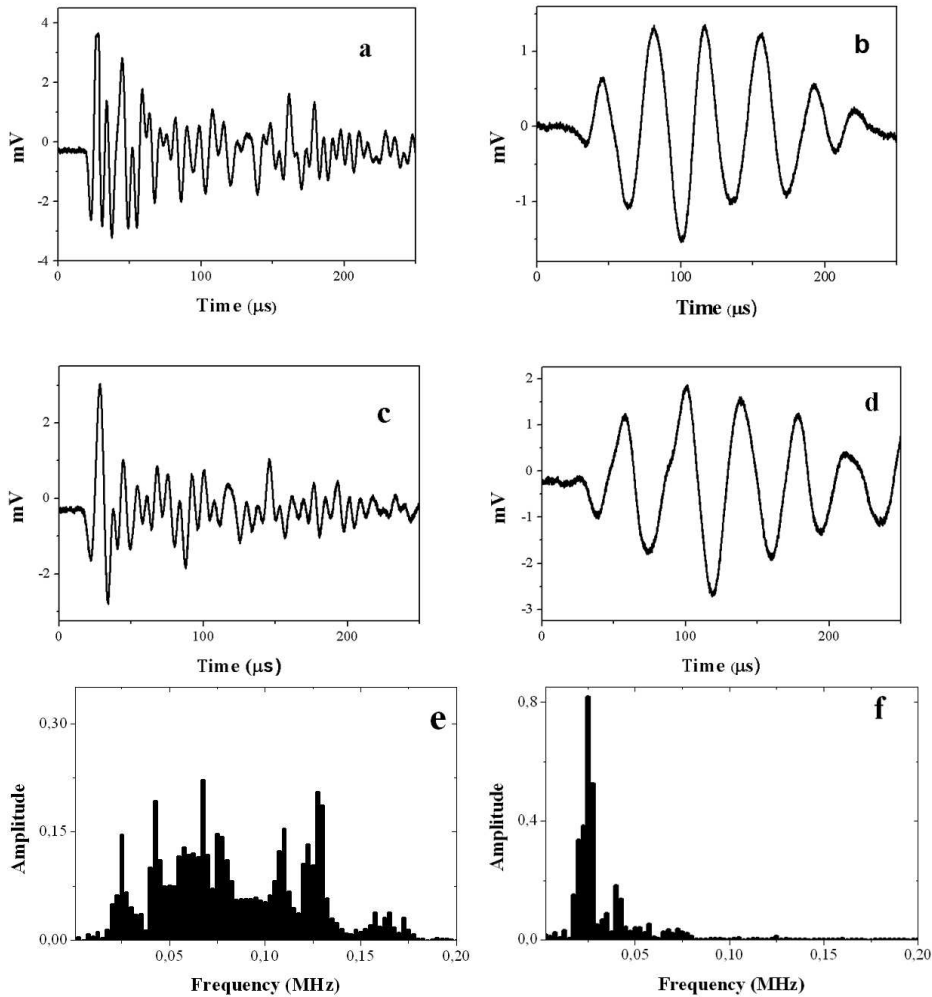}
\caption{Examples of the detected UAE related to cavitation events in corn stem. (a) Type 1 broadband frequency emission signals detected at room light. (b) Type 2 low frequency signals detected at room light. (c) Type 1 signals detected with the laser impinging near the detector. (d) Type 2 signals detected with the laser impinging far from the detector. (e) and (f) Frequency spectra of type 1 and type 2  signals detected with laser. See the similarity of the signals produced with laser and those detected at room light.} \label{figure2}
\end{center}
\end{figure*} 

\section{Results}

\subsection{Measurements of spontaneous UAE}

In the first experiment (\textit{Experiment 1}), a transducer was attached to the bare stem on an internode with 5--7 developed leaves above it. UAE were monitored in the dark (D), under room light (RL) and under incandescent lamp (IL) illumination. The experiment was performed with several plants changing the sequence conditions of light (D--RL--IL; RL--D--IL, etc.). We registered no emissions in the dark. In experiments 2--3 h long, a rate of 1.15 $\pm$ 0.09 emissions min$^{-1}$ were detected when the plant was transpiring under ambient light. This rate was increased to 1.45 $\pm$ 0.15 emissions min$^{-1}$ when transpiration was stimulated with an incandescent lamp. The change in the rate of emissions between RL and IL took place less than a minute after turning the lamp on or off. When two transducers were attached to the bare stem at the same height but in different radial positions [Fig. 1(B)], the rate of emission and the type of signals (see below) were the same for both detectors.

 In a second set of experiments performed under room light, and each lasting ca. 2 h (\textit{Experiment 2}), two transducers were attached to the bare stem at different heights. Both transducers registered UAE, but the rate of emissions was dependent on the transducer position: near the leaves was higher than closer to the plant base. For instance, when T1 was located at 14 cm from the plant base and T2 at 35--40 cm, no signals were detected by T1, while 0.3--3.5 emissions min$^{-1}$ were detected by T2. The amplitude of the signals detected by each transducer was registered as a function of time, and the UAE were classified by their form and main frequency. Two types of signals were identified: those who have a broad band of frequencies up to 0.2 MHz, named type 1, [Fig. 2(a)], and low frequency signals, with values below 0.075 MHz, named type 2, [Fig. 2(b)].

\subsection{Laser induced UAE}

With the aim of developing a method to induce UAE in a controlled way, in the next series of experiments (\textit{Experiment 3}) we impinged a laser beam at a point on a corn stem with a transducer attached on the opposite side [Fig. 1(C)]. We started measuring UAE in the dark, and without laser irradiation. Under these conditions, no UAE were detected. Then, again in the dark, we irradiated the stem with the He--Ne red laser, but even at its maximum power, no UAE were detected. After that, the CW Ar-ion laser was tested at different wavelengths and powers. We found that with powers up to 600 mW, only the blue line at 488 nm produced results. Under these conditions, when the laser was turned on, acoustic signals were registered and when it was turned off, the rate of emission decayed and disappeared after a few seconds (Fig. 3).

\begin{figure}[th]
\begin{center}
\includegraphics[width=0.8\columnwidth]{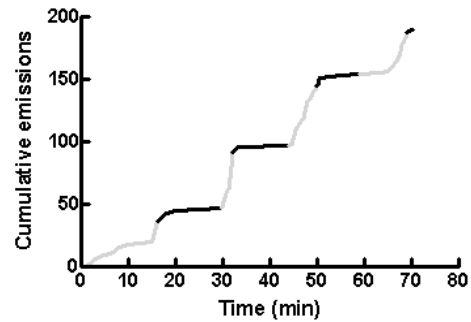}
\caption{Laser induced UAE in corn stem. The beam of a CW Ar ion laser (600 mW) at 488 nm impinges on the stem opposite to the transducer. Grey line: the laser is on. Black line: the laser is off.} \label{figure3}
\end{center}
\end{figure} 

This sequence (switching the laser on and off, always impinging on the same point of the stem) was repeated with the same qualitative results, although the rate of UAE decreased with every cycle (in Fig. 3 compare the slope of the sequence starting at minute 40 with the one starting at minute 55). Even when the rate of emissions in different plants encompassed a wide range (ca. 2--17 emissions min$^{-1}$ with the laser on), the same pattern always held (i.e., emissions when laser is on, and no emissions a few seconds after the laser is off). The same behavior was observed when the laser impinged at a right angle from the transducer axis. 

\begin{figure}[th]
\begin{center}
\includegraphics[width=0.7\columnwidth]{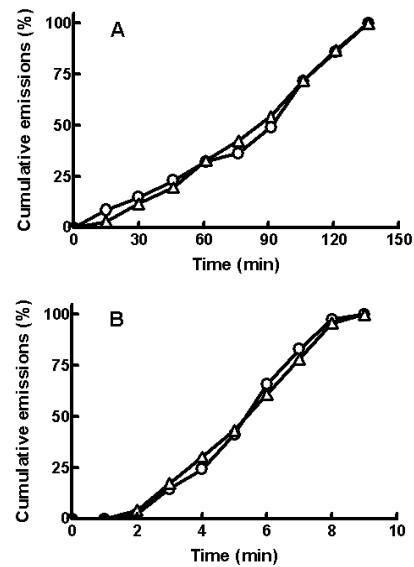}
\caption{(A) Spontaneous and (B) laser induced UAE in corn stems measured simultaneously with two transducers attached at the same height of the stem. In (B) the laser impinged between both transducers. Open triangles: transducer T1. Open circles: transducer T2.} \label{figure4}
\end{center}
\end{figure}

The signals were classified according to their form and frequencies. Figure 2(c) shows a typical signal generated by the laser in this experiment. As it can be seen, these signals are similar to the broadband frequency signals [the type 1 shown in Fig. 2(a)] measured in experiment 2 with room light. 

With the aim of comparing spontaneous and laser-induced UAE, we attached two transducers to the bare stem on opposite sides, at the same height [\textit{Experiment 4}, Fig. 1(B)]. We first registered acoustic emissions detected simultaneously by both transducers at room light, without laser irradiation [Fig. 4(A)]. After that, in the dark, we measured the UAE generated after impinging the CW laser between both detectors, in a direction perpendicular to their axes [Fig. 4(B)]. The characteristic signals observed in both cases were type 1 signals (broadband frequency signals).

Then, we proceeded to study how the distance between detector and source modified the rate and shape of the UAE (\textit{Experiment 5}). Two transducers were attached to the bare stem: one at 8.5 cm (T1) and the other at 12.5 cm (T2) from the base. The CW laser beam impinged on different points of the stem. Points \textbf{a} and \textbf{b} were at the same height of T1 and T2, respectively, but at the opposite side; point \textbf{c} was between T1 and T2 [Fig. 1(D)].

\begin{figure}[th]
\begin{center}
\includegraphics[width=0.7\columnwidth]{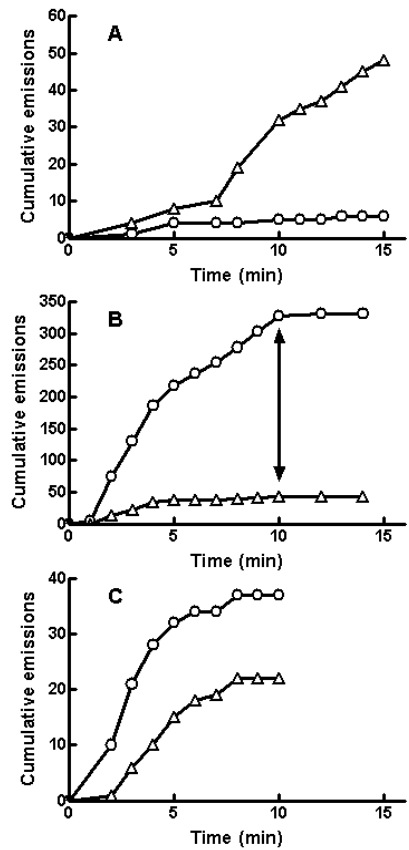}
\caption{Laser induced UAE as a function of the transducer position. Two PZT transducers were attached to the stem at two heights as shown in Fig. 1(D). Open triangles: transducer T1. Open circles: transducer T2. (A) The Laser beam impinged near T1. (B) The Laser beam impinged near T2. (C) The Laser beam impinged between T1 and T2. The arrow in (B) indicates the laser was off.} \label{figure5}
\end{center}
\end{figure} 

When the laser beam impinged on \textbf{a}, both transducers detected UAE. Broadband frequency signals (type 1) were observed with T1 and low frequency signals (type 2) were observed with T2. Figure 2(d) shows an example of type 2 signals generated with laser. As it can be seen, these signals are similar to those detected at room light [Fig. 2(b)].

 Besides, the number of emissions detected by T1 was higher than that detected by T2 [Fig. 5(A)]. When the laser beam impinged on \textbf{b}, once again, both transducers detected UAE. In this case, T2 detected type 1 signals while T1 detected type 2 signals, and the number of emissions detected by T2 was higher than that detected by T1 [Fig. 5(B)]. When the laser beam impinged on \textbf{c}, both transducers simultaneously detected UAE of low frequency similar to those described as type 2 [Fig. 5(C)]. 

\section{Discussion and conclusions}

Experiments 1 and 2 show that the spontaneous UAE can be attributed to natural cavitation events occurring in the xylem vessels of the corn stem: no emissions were observed when the plant was in the dark. Around 1 emission min$^{-1}$ was detected under room light, and a rate ca. 25\% higher under the lamp. This behavior is in agreement with the cohesion-tension theory and current plant cavitation models. As transpiration rate increases, xylem tension rises and cavitation events are expected to increase, as it happens in our experiments. Besides, the UAE signals registered (Fig. 2) were very similar to those described \cite{12}. These authors demonstrated that these kinds of emissions are strongly related to cavitation events \cite{12,13,26}.

In the transpiring plant, the tension developed in the water stream generates a metastable equilibrium. When liquid water is subjected to a sufficiently low pressure, this equilibrium can be broken, and form a cavity. This initial stage of the cavitation phenomenon is termed cavitation inception. When the plant is in the dark, water in the xylem is slightly under tension at a pressure value close to atmospheric. Under these conditions, the local pressure does not fall enough, compared to the saturated vapor pressure, to produce cavitation inception. 

As the CW laser impinges on the stem, this absorbs light and release energy to the xylem, heating it. This extra energy allows the phase change to gas in the water column, triggering cavitation inception. In this sense, the physical process of cavitation inception is similar to boiling, the major difference being the thermodynamic path which precedes the formation of the vapor. We found that UAE generated using a CW laser are of the same kind of those registered on transpiring plants. We can conclude that this method allows, for the first time, to induce cavitation events in xylem in a controlled and reproducible way.

Regarding the mechanisms of UAE generation, either natural or laser-induced, previous work has clearly shown that once cavitation inception is produced, embolism of the xylem immediately takes place. This means that the formed cavity remains, and there is no collapse of the void in the water column (as would occur in the so called inertial cavitation). Then, UAE generation can be produced by an oscillating source activated by the rupture of the water column. As mentioned in the introduction, Tyree \textit{et al.} \cite{12} proposed four possible UAE sources. The first one, oscillation of hydrogen bonds in water after tension release, seems unlikely because of its very low magnitude, undetectable by the kind of transducers we used. The second one, oscillations caused by a ``snap back'' of vessel walls, is also unlikely because of the rigidity of the xylem, and especially hard to explain under laser-induced cavitation inception in the dark, when xylem tension was nil or very small. The third one, torus aspiration, is impossible in our case because of the absence of these structures in corn. Finally, the fourth one, structural failure in the sapwood, was elegantly rejected by Tyree himself \cite{13}, who exposed xylem to pressure and detected a different kind of emission.

We postulate that another possible source of the UAE must be taken into account. It is the local oscillation of the liquid--gas interface of the water column produced by the expansion and compression of the formed cavity, i.e., the stress wave generated by rapid bonding energy release. During cavitation inception, after the cavity expands, it is expected to be compressed almost immediately by the water column.  This ``cavity oscillation'' starts as a high frequency burst produced by the disruptive effect of the cavitation inception. As a consequence, ultrasonic acoustic signals are produced.

In order for cavitation inception to occur, the cavitation ``bubbles'' generally need a surface on which they can nucleate. This could be provided by impurities in the liquid or the xylem walls, or by small undissolved micro-bubbles within the water, but most likely by air seeding through pit membranes \cite{4}. These act as capillary valves that allow or prevent air seeding by adjusting local curvatures and interface positions \cite{27}. Air seeding induced by the heating at pit membranes under CW laser irradiation should also be taken into account as an initial stage in laser induced cavitation inception. 

The CW laser-induced cavitation opens the opportunity to study embolism in plants in a controlled manner. It also has the advantage of tracing the source, allowing the characterization of the signals and studying their propagation. By directing the laser beam to one point in the stem and recording acoustic emissions at different distances, we found that when cavitation was produced near the transducer, broadband frequency emissions were registered. But, if the transducer was installed further away, the rate and  frequency of the emissions decreased with the distance to the cavitation source. This means that during signal propagation, absorption by the tissue takes place (rate decay) as well as frequency filtering. Figures 2(e) and 2(f) show the frequency spectra of type 1 and type 2 signals. When comparing these figures, the frequency filtering effect is evident.

Our results confirm the hypothesis by Ritman and Milburn \cite{28}, who proposed that cavitation of xylem sap generally results in the production of a broadband acoustic emission with lower cut-off frequency determined by the dimensions of the resonating element. The larger a conduit dimension, the lower the frequency of its major resonance. Thus, small cavitating elements, such as corn stem xylem, are expected to produce acoustic signals with a broadband frequency spectrum. Our results can also explain the observations by Tyree and Dixon \cite{12} who found and classified UAE of different frequencies (between 0.1 and 1 MHz). According to our experiments, the different signals would be generated by cavitation events produced in different regions of the stem. Broad band frequency signals would come from near the transducer while low frequency signals would come from regions far from to the transducer.

According to these results, one might use the waveform of the emissions to determine the location of each cavitation event. In that case, a whole new field would be opened in the study of hydraulic architecture of plants.

\begin{acknowledgements}
The authors are indebted to Dr. H. F. Ranea Sandoval of FCE-UNCBA-Tandil-Argentina, Professor Silvia E. Braslavsky from Max-Planck-Institut f\"{u}r Bioanorganische Chemie M\"{u}lheim an der Ruhr, Germany and Dr. J. Alvarado-Gil from CINVESTAV-Unidad, Merida, Merida, Mexico for fruitful comments and suggestions. This work was partially supported by ANPCyT, UNCPBA, UBA and UNLP. G.M.B. is member of the Carrera del Investigador Cient\' {\i}fico CIC-BA, and R.J.F. of CONICET.
\end{acknowledgements}

\end{document}